\documentstyle[11pt,newpasp,twoside]{article}
\def\edcomment#1{\iffalse\marginpar{\raggedright\sl#1\/}\else\relax\fi}
\reversemarginpar

\begin{document}
\title{The Formation and Evolution of Star Clusters and Galaxies}
\author{Stephen E. Zepf}
\affil{Dept. of Physics and Astronomy, Michigan State University, 
East Lansing, MI 48824, USA}

\begin{abstract}
	This paper addresses the questions of what we have learned
about how and when dense star clusters form, and what studies
of star clusters have revealed about galaxy formation and evolution.
One important observation is that globular clusters are observed to form
in galaxy mergers and starbursts in the local universe, which
both provides constraints on models of globular cluster formation, 
and suggests that similar physical conditions existed when most
early-type galaxies and their globular clusters formed in the past. 
A second important observation is that globular cluster systems
typically have bimodal color distributions. This was predicted
by merger models, and indicates an episodic formation history for
elliptical galaxies. A third and very recent result is the discovery 
of large populations of intermediate age globular
clusters in several elliptical galaxies through the use
of optical to near-infrared colors. These provide an
important link between young cluster systems observed in
starbursts and mergers and old cluster systems. This
continuum of ages of the metal-rich globular cluster
systems also indicates that there is no special
age or epoch for the formation of the metal-rich globular 
clusters, which comprise about half of the cluster population.
The paper concludes with a brief discussion of
recent results on the globular cluster -
low-mass X-ray binary connection.

\end{abstract}

\section{Globular Cluster Formation}

A natural starting point for the discussion of
galaxies and globular cluster systems is the formation of
globular clusters (GCs). Any model of the formation of globular
cluster systems and their host galaxies that does {\it not}
include a consideration of how the globular clusters themselves
form is necessarily incomplete. Fortunately, nature has
provided us nearby examples of GC formation,
most dramatically revealed in HST images (e.g.\ Whitmore et al.\ 1999, 
Zepf et al.\ 1999).
The formation of globular clusters in galaxy mergers was predicted 
by Ashman \& Zepf (1992) and Schweizer (1987) 
and then confirmed by many subsequent HST observations. Moreover, 
as discussed at this meeting by Bruce Elmegreen, these observations 
are providing an important testing ground for theoretical work on 
GC formation (e.g.\ Elmegreen 2002, Ashman \& Zepf 2001).

\section{Key Properties of Globular Cluster Systems}

	A second key point is the bimodality typical
of the color distributions of early-type galaxies.
The bimodality of early-type galaxy gobular cluster
systems was first discovered about ten years ago
(Zepf \& Ashman 1993), and has now been confirmed 
to be typical by extensive studies of archival
HST data (Kundu \& Whitmore 2001, Larsen et al.\ 2001).
The data show that roughly half of the GCs are blue 
(metal-poor) and half are red (metal-rich, i.e.\ very roughly 
solar or somewehat less). There is little or no trend 
of the red/blue ratio with galaxy luminosity or globular cluster number
(e.g.\ Rhode \& Zepf 2003).

	The observed bimodality in the cluster systems
of elliptical galaxies has important implications for
the formation of globular clusters and galaxies. Firstly,
bimodality was predicted by Ashman \& Zepf (1992) to result
from elliptical galaxy formation by mergers of disk galaxies, 
mostly at $z \ga 1$, and so the observation was 
in agreement with an earlier theoretical prediction.
Moreover, regardless of the specific model,
bimodality requires an episodic formation history for
elliptical galaxies, and not a single formation event. 
Secondly, most of these events also have to occur at least 
$\sim 8$ Gyr ago in order for the metal-rich GCs to become 
red enough to produce the color bimodality from the metallicity 
bimodality. However, even 
if most of the major formation events take place at $z \ga 1$,
some will take place more recently, and to make the picture
complete, we should be able to identify and characterize these.
This has recently been achieved, as discussed in the following
section.

\section{Intermediate Age Globular Cluster Systems}

	The discovery of intermediate age globular cluster
systems in elliptical galaxies is important because it links 
the young systems observed in mergers to the traditional old 
GC systems. A key observational advance has been the use
of optical to near-infrared colors to break the age-metallicity
degeneracy to identify intermediate age, $\sim$ solar metallicity 
globular clusters (which are effectively indistinguishable from 
lower metallicity, older clusters in optical colors alone).

	The breakthrough in identifying intermediate age
clusters came in two ways. Puzia et al.\ (2002) discovered
a large population of intermediate age globular clusters
in the otherwise fairly normal elliptical galaxy NGC~4365.
Subsequent spectroscopy of a subset of the Puzia et al.
clusters confirmed the effectiveness of their optical
to near-infrared technique (Larsen et al.\ 2003).
While Puzia et al.\ (2002) demonstrated that some quiescent
ellipticals had a major formation event in their not so distant
past, Goudfrooij et al.\ used similar techniques to identify
a major intermediate-age globular cluster population in the 
disturbed galaxy NGC~1316. More recently, Hempel et al.\ (2003)
found a significant intermediate-age cluster population in
NGC~5846. Both Puzia et al.\  and Hempel et al.\ 
also find galaxies without intermediate age populations,
as expected since the common bimodal systems are probably
not intermediate-age. However, the presence of intermediate
age systems shows that the age distribution of cluster systems
is continuous from very young to very old, with all ages
present. 
The formation of these clusters traces out the starburst
histories of their host ellipticals, which appears to
be complex and to occur over a significant range of times.

\section{Globular Cluster - Low-Mass X-Ray Binary Connection}
	Since this is a dynamics session, and GCs are fertile
ground for the study of stellar dynamics, it is relevant
to note that the dynamical interactions between stars
play an important role in understanding X-ray emission
from elliptical galaxies. This is because roughly half of 
all X-ray binaries seen in Chandra images are located
in GCs (see upcoming review by Verbunt \& Lewin 2003
and references therein).
Moreover, the dynamical formation of X-ray binaries in GCs
may be the dominant process in early-type galaxies, since
sources not now in GCs may have been formed there and
ejected or been in clusters that were disrupted
(e.g.\ Maccarone, Kundu, \& Zepf 2003).
The study of the GCs hosting low-mass X-ray binaries
also provides constraints on the physics of X-ray binary
formation and evolution. In particular, there is a strong metallicity
dependence and no strong age dependence (Kundu et al.\ 2003
and references therein).

\acknowledgments
	I thank my many collaborators on
the projects described here, and also my postdocs and
graduate students
Arunav Kundu, Katherine Rhode, Enrico Vesperini,
Gilles Bergond, and Chris Waters. Financial support is
acknowledged  from LSTA grant NAG5-11319
and Chandra and HST programs. 

\references

Ashman, K.M., \& Zepf, S.E. 1992, ApJ, 384, 50


Ashman, K.M., \& Zepf, S.E. 2001, AJ, 122, 1888

Elmegreen, B.G. 2002, ApJ, 577, 206

Goudfrooij, P., et al.\ 2001, MNRAS, 322, 643

Hempel, M., et al.\ 2003, A\&A, 405, 487

Kundu, A., \& Whitmore, B.C. 2001, AJ, 121, 2950

Kundu, A., Maccarone, T.J., Zepf, S.E., \& Puzia, T.H. 2003, ApJL, 
589, 81

Larsen, S.S., et al.\ 2001, AJ, 121, 2974

Larsen, S.S., et al.\ 2003, ApJ, 585, 767

Maccarone, T.J., Kundu, A., Zepf, S.E. 2003, ApJ, 586, 814

Puzia, T.H., Zepf, S.E., Kissler-Patig, M., Hilker, M., Minniti, D., \& 
Goudfrooij, P. 2002, A\&A, 391, 453

Rhode, K.L., \& Zepf, S.E. 2003, AJ, submitted

Schweizer, F. 1987, in Nearly Normal Galaxies, ed.\ S.M. Faber
(Berlin: Springer), 18



Whitmore, B.C., Zhang, Q., Leitherer, C. Fall, S.M.,
Schweizer, F., \& Miller, B.W. 1999, AJ,  118, 1551

Zepf, S.E., \& Ashman, K.M. 1993, MNRAS, 264, 611

Zepf, S.E., Ashman, K.M., English, J., Freeman, K.C., \& Sharples, R.M.
1999, AJ, 118, 752

\end{document}